Eddington and Uncertainty

Ian T. Durham[*]

Arthur Stanley Eddington (1882-1944) is acknowledged to be one of the greatest astrophysicists of the twentieth century, yet his reputation suffered in the 1930s when he embarked on a quest to develop a unified theory of gravity and quantum mechanics. His attempt ultimately proved to be fruitless and was regarded by many physicists as misguided. I will show, however, that Eddington's work was not so outlandish. His theory applied quantum-mechanical uncertainty to the reference frames of relativity and actually foreshadowed several later results. His philosophy regarding determinism and uncertainty also was quite orthodox at the time. I first review Eddington's life and philosophy and then discuss his work within the context of his search for a theory of quantum gravity.

*Key words*: Arthur Stanley Eddington; Edward Arthur Milne; Joseph Larmor; history of cosmology; relativity; quantum mechanics; quantum gravity.

**Introduction**

Albert Einstein's theory of general relativity stands among the most significant developments in the history of modern cosmology. Einstein described gravity as a consequence of geometry. Subsequently, physicists attempted to link gravity to the other force of nature known at that time, electromagnetism, in an attempt to unify physics. In 1921 the German theoretical physicist Theodor Kaluza (1885–1954) was the first to attempt to unify relativity and electromagnetism by extending Einstein's field equations to five dimensions,[1] an idea that was improved upon in 1926 by the Swedish theoretical physicist Oskar Klein (1894–1977).[2] Interest in unification then waned with the creation of quantum mechanics and its blow against determinism. Some physicists, however, did not abandon the quest for unification; Einstein, in fact, devoted the last thirty years of his life to it.

[*] Ian T. Durham is Visiting Assistant Professor of Physics at Simmons College in Boston and a doctoral candidate in the School of Mathematics and Statistics at the University of St. Andrews in Scotland.

Unification is widely regarded today as the Holy Grail of physics. Physicists have successfully wedded the strong, weak, and electromagnetic forces, but their marriage to gravity remains unconsumated. Wedding gravity and quantum theory is at the heart of this quest, and theories of quantum gravity now have been at the forefront of research in physics for nearly forty years. But attempts at unification never died out entirely in the years immediately following Kaluza's and Klein's work. Paul Adrien Maurice Dirac (1902-1984) made one of the first attempts in 1928 with his relativistic equation for the electron.[3] Arthur Stanley Eddington (1888-1944), disappointed that Dirac's equation did not appear in tensor form,[†] sought to reformulate Dirac's work in 1929-1930 to put quantum theory into the language of relativity.[4] Thus began a grand, though often unfruitful series of cosmological theories developed by Eddington, Dirac, and Edward Arthur Milne (1896-1950).

Eddington's work, which rested on the premise that quantum mechanics and relativity could be united under a common framework, centered on the idea of uncertainty in the reference frames of relativity as we will see. He was convinced that the introduction of uncertainty into physics heralded such a monumental change that everyone had to consider its philosophical implications. He was firmly committed to the Copenhagen interpretation of quantum mechanics; to him uncertainty was inherent and inescapable. He denigrated the opposing Einstein-Podolsky-Rosen (EPR) interpretation, saying that any scientist who accepted the idea of hidden variables as an explanation of indeterminacy "wants shaking up and waking."[5] Eddington saw the fundamental indeterminacy in the quantum world as the foundation on which to build a unified theory of physics.

---

[†] Charles Galton Darwin (1887-1962) was the first to note that Dirac's equation was not in tensor form; see C.G. Darwin, "The Wave Equation of the Electron," *Proceedings of the Royal Society* [A] **118** (1928), 654-680.

Eddington's work on uncertainty, though not completely successful, actually foreshadowed some later developments in physics, including the need for a quantum-mechanical standard of length. This led him to the ultimate framework of later versions of his theory, which held that physical events depend solely on dimensionless numbers. His idea was taken up later by Dirac in proposing his Large Numbers Hypothesis,[6] which was based on this premise.

Other aspects of Eddington's work hinted at some of the underlying principles of modern quantum field theory and string theory. Kaluza-Klein theory is an active area of research today, and Eddington made some suggestions that are close to modern ideas. Eddington was a consummate mathematician; even his earliest works were intensely mathematical. But he also had a knack for interpreting mathematics and sorting out their meaning. It was precisely the mathematics underlying his physical theories that launched his attempt to unify relativity and quantum mechanics: His dissatisfaction with Dirac's relativistic theory of the electron launched his work on uncertainty, which ceased with his death in 1944 and with the posthumous publication of his monumental *Fundamental Theory* in 1946.[7]

**The Quiet Genius**

> He came from a family of quiet people – long and sincerely devoted to the principles of the Society of Friends – and, apart from the honors brought him by his work, his life was uneventful. [8]

So wrote Henry Norris Russell (1877–1957), whose characterization of Eddington's life may seem sad or even unkind, but Eddington himself probably would have agreed with it. He lived simply, yet he was a complex man. He was devoted to those he cared for, and was passionate about everything that was important to him. By many people's standards, his life *was* uneventful, but it was filled with simple pleasures. To many his life very likely was secretly enviable.

Arthur Stanley Eddington was born on December 28, 1882, in Kendal, Westmoreland, England. He was the second child and only son of Arthur Henry Eddington, who was the headmaster of the Stramongate School, the Society of Friends school where the chemist John Dalton (1766–1844) once taught.[9] The Quaker tradition was omnipresent throughout Eddington's life and brought him the inner serenity to focus intensely on scientific problems. It embodies a philosophy of peace and inner harmony, not unlike some Eastern philosophies; at the same time, it was a source of tension for him at times. During the Great War of 1914-1918, conscientious objectors were placed in camps. Quakers usually became conscientious objectors, and Eddington had friends who were sent to pick potatoes in agricultural camps.[10] The mathematical physicist Joseph Larmor (1857–1942) tried to utilize contacts at the British Home Office to have Eddington deferred on the grounds that it was not in the national interest to have a distinguished scientist in the Army.[‡] Eddington added a postscript to Larmor's letter to the Home Office, saying that if he was denied an exemption on the grounds of his usefulness to British science, he would claim conscientious-objector status. "Larmor and others were very much piqued."[11] Eddington's action thus led to a short, heated exchange of letters with Larmor, who seems to have insinuated in one (which I have been unable to locate) that conscientious objectors held pro-German views. Eddington's response (which has survived in Larmor's correspondence) argues strongly and passionately against this view but is courteous throughout.[12] Such passion and courtesy was characteristic of Eddington and surfaces often in his personal correspondence and scientific papers.[13] Despite his heated exchange with Larmor, he did not hold a grudge; the two later were on friendly terms. This pattern was repeated with Milne, one of his strongest professional antagonists and one of his closest personal friends.

Eddington was a precocious and intelligent child; he attempted to count the words in the Bible and mastered the 24 x 24 multiplication table before he could read.[14] He obtained a three-inch telescope a bit later and immediately turned it to the skies, thus beginning his lifelong study of the heavens.[15] His schooling was a whirlwind of success beginning with the Brynmelyn School in Weston-super-Mare (1893-1898), where his family had moved shortly after the untimely death of his father in 1884. He then attended Owens College, Manchester (1898-1902), circumventing rules prohibiting those under the age of 16 from entering. His professors at Owens included mathematician Horace Lamb (1849-1934) and physicist Arthur Schuster (1851-1934). In 1902 he transferred to Trinity College, Cambridge, on a scholarship. In 1904, in his second year, he became Senior Wrangler in Cambridge's rigorous Mathematical Tripos Examination, an unprecedented achievement.§ In 1907, he was awarded the Smith's Prize and elected as a Fellow of Trinity College. Earlier, in 1905, he spent a term working in the Cavendish Laboratory where he very nearly made a career for himself in physics. Had he become a physicist, he might not have learned valuable lessons that later led him to his *Fundamental Theory*. Instead, in 1906, he was appointed as Chief Assistant at the Royal Observatory, Greenwich, to succeed Sir Frank Dyson (1868-1939) who moved to Edinburgh as Astronomer Royal for Scotland. Once again, in 1909, he was tempted by physics but turned down Schuster's offer of a position in physics at Manchester because, as he explained in a letter to Schuster, he preferred observational work.[16] He returned to physics later in his career, as we shall see, but his philosophical outlook was shaped mostly by his astronomical work.

---

[‡] Physicist Henry G. J. Moseley (1887–1915) had been killed at Gallipoli, as Larmor reminded the Home Office. Moseley had worked with C.G. Darwin and Ernest Rutherford in Manchester and had used X-ray spectra to study atomic structure, which laid the groundwork for ordering the elements in the periodic table.

In 1913, at the young age of 31, Eddington succeeded Sir George Darwin (1845-1912), son of evolutionist Charles Darwin (1809-1882), as the Plumian Professor of Astronomy and Experimental Philosophy at Cambridge. The following year he also succeeded Sir Robert Ball (1840-1913) as Director of the University Observatory and was elected as a Fellow of the Royal Society. He also was elected to the Royal Astronomical Society in 1906 and served as its Secretary from 1912 to 1917, as its President from 1921 to 1923 (figure 1), and as its Foreign Secretary from 1933 until his death.  He received the Society's Gold Medal in 1924 and a Royal Medal in 1928.  The Astronomical Society of the Pacific awarded him its Bruce Medal in 1924.  He became President of the International Astronomical Union in 1938.  As a British subject, he received two of Britain's highest honors: He was knighted in 1930 and received the Order of Merit in 1938.  His knighthood status was that of Knight Bachelor; he never married.

---

§ The Senior Wrangler was the student who gained the highest marks on the Mathematical Tripos Examination; next came the Second Wrangler, then the Third, and so on. This system was replaced in 1909 with a purely alphabetical placement list.

This is telling.  Eddington never entered into anything lightly in his science, and his bachelorhood indicates that this was true in his private life as well.  But human relationships were important to him.  W.M. Smart maintains that his shy, quiet fortitude was more evident in his later years, while in his early years in Greenwich he was a member of the Observatory Hockey team, not a sport given to shyness.[17]  He also was an accomplished golfer and regular swimmer, often plying the River Cam.  He had a lifelong interest in cycling and rather wistfully recounts his cycling trips in letters to friends.[18]  He seems to have drawn inward around the time that his science turned from observational and inductive astronomy to theoretical and deductive physics, a point to which I will return.

Eddington was a master at language, in English as well as the classics.  He was well versed in literature and wrote poems for his own amusement that were considered meritorious by those lucky enough to steal a glimpse of them.  He was a fan of Lewis Carroll (1832-1898), himself a mathematician, and often made up grammatically correct sentences in Carroll's style that made no sense.  Subramanyan Chandrasekher (1910-1995) gives an example directly told to him by Eddington: "To stand by the hedge and sound like a turnip."[19]  Some of his more lighthearted fare was actually published, for example:

> There once was a brainy baboon,
>
> Who always breathed down a bassoon,
>
> For he said, "It appears
>
> That in billions of years
>
> I shall certainly hit on a tune."[20]

Harold Spencer Jones (1890-1960) and Edmund T. Whittaker (1873-1956) recalled his "retentive memory for the apposite quotation" and his fondness for Shakespeare, having been a member of The Elizabethans, a small private society devoted to The Bard at Greenwich.[21] He was addicted to solving crossword puzzles in *The Times* and the *New Statesman and Nation* and rarely took more than five minutes per puzzle.[22] He displayed his linguistic skill in several popular books on science and philosophy. He also enjoyed writing mathematical puzzles, the most famous of which – the Zoo Puzzle[**] – still circulates among puzzle enthusiasts.

Eddington, in sum, was a highly complex man whose philosophical commitments permeated his life and his science.

**Explicit Philosophy**

There is a hidden meaning in the above doggerel. The bassoon-playing baboon is banking on the laws of probability to assure him that he will eventually "hit on a tune." Probability constituted the mathematical backbone of Eddington's philosophy, which he constructed on the concept of uncertainty, the cornerstone of his attempt to unify relativity and quantum theory (see below).

---

[**] The Zoo puzzle is a problem in combinatorics written in the style and using the characters of Lewis Carroll.

Since the mathematical underpinnings of uncertainty are rooted in probability theory, it was natural for Eddington to grapple with the nuances of probability. His interest in probability began in the 1920s when he worked on unifying gravity and electromagnetism. He lectured regularly on the subject; for example, when he published an expanded version of his 1934 Messenger Lectures at Cornell University as *New Pathways in Science*, he devoted entire chapters to the decline of determinism, indeterminacy and quantum theory, and probability.[23] In them he refers numerous times to Pierre-Simon Laplace (1749-1827), a deductivist whose work in mathematics, including probability theory, was based on logical reasoning. Since Eddington too believed that a truly unified theory of the universe should be derivable solely from logical reasoning, he too was a deductivist and sympathetic to the work of Laplace.

Laplace was at the center of a heated debate in Britain in the 1820s when an attempt was made to introduce the teaching of probability theory into the curriculum of the Universities of Cambridge and Oxford, which were then still sectarian (in contrast to the secular University of London). In particular, several religiously fervent dons at Cambridge, including William Whewell (1794–1866), argued against introducing probability theory into the curriculum, opposing the teachings of Continental deductivists such as Laplace, Jean D'Alambert (1717-1783), Alexis Claude Clairault (1713-1765), Joseph-Louis Lagrange (1736-1813), and Leonhard Euler (1707-1783) on the grounds that probability theory sought to answer questions better left to the Divine. In his *New Pathways in Science*, Eddington relates the story of Marie Jean Antoine Nicolas de Caritat (1743-1794), the Marquis de Condorcet, another French probabilist, who attempted to apply probability theory to the fairness of judges and took his own life after calculating his odds of acquittal by the Revolutionary tribunal.[24] Whewell favored inductive science, which was experimental in nature and thus supposedly more supportive of religious beliefs; deductive science was too mechanistic. Joan Richards has noted that "a religion that rested on evidence attested to by personal experience and conviction had no standing in probabilistic discourse."[25]

Eddington, as we have seen, began his career as an observational astronomer, and his extensive observational work possibly led him down the deductivist path. By today's standards, observational astronomy in Eddington's day was a highly inexact science. Eddington, the consummate mathematician, attempted to formulate exact mathematical theories to match astronomical observations, as in the case of his study of Comet Morehouse in 1909 when he attempted to fit three-dimensional paraboloids to the envelopes of the comet from fuzzy, two-dimensional photographic plates, finding considerable room for error.[26] Similarly, in his famous solar-eclipse expedition of 1919, it turned out that only one of his numerous observations was good enough to support Einstein's general theory of relativity.[27] The consistent inadequacies of observational techniques, particularly when compared to the rigor of mathematics, must have convinced Eddington psychologically that some level of uncertainty was inherent in any observation. Werner Heisenberg's uncertainty principle thus must have offered Eddington exactly what he was seeking as a philosophical foundation for his work. As we will see, he extended Heisenberg's uncertainty principle from microscopic to macroscopic phenomena by introducing uncertainty into the reference frame of any observation, and he also grappled deeply with the problem of how to define standards of measurement like the meter. Eddington thus became a deductivist who sought to determine everything from logical reasoning. Ironically, Whewell's inductivism incorporated what he called "fundamental ideas" (could Eddington have gotten the title of his book from Whewell?) that were supported by observation but were found by "thinking properly."

**Implicit Philosophy**

The distinction between determinacy and indeterminacy in physics can be described in terms of inductive and deductive reasoning, with inductivists falling into the determinacy camp and deductivists falling into the indeterminacy camp. We can turn to the definitions of inductive and deductive reasoning as given in *Warriner's English Grammar and Composition*.[28] Inductive reasoning starts from a set of observations and draws a generalization from them. Deductive reasoning starts with a generalization and draws conclusions from it. In early Victorian science, probability was established as a form of deductive reasoning. Its conclusions were derived from mathematical first principles and did not rely on experimental evidence. Experimental science, by contrast, was established as a form of inductive reasoning.[29]

Eddington's early career was marked by the inductivism inherent in observational astronomy. Perhaps motivated by its inadequacies at the time, he turned later to the more deductivist science of cosmology. But deductivism entails a definite problem in relation to science. As Bertrand Russell (1872-1970) wrote in his *Introduction to Mathematical Philosophy*:

> Since all terms that are defined are defined by means of other terms, it is clear that human knowledge must always be content to accept some terms as intelligible without definition, in order to have a starting-point for definition. It is not clear that there must be terms which are *incapable* of definition: it is possible that, however far back we go in defining, we always *might* go further still.[30]

This sums up Eddington's philosophy as he grappled with uncertainty. Uncertainty arose because there was never a suitably defined starting point in measurement. Eddington's deductivism also finds a parallel in another passage in Russell where he notes that Gottlob Frege (1848-1925) "first succeeded in 'logicising' mathematics, *i.e.* in reducing to logic the arithmetical notions which his predecessors had shown to be sufficient for mathematics."[31] Eddington attempted to "logicise" *physics* by reducing it to a logical set of arithmetical notions that had been shown to be sufficient for *physics* (or by reducing physics to mathematics in those cases where mathematics can be applied in physics). The problem here, of course, is how to deduce a definite conclusion from an indefinite starting point. The major philosophical flaw in Eddington's approach, then, was his assumption that a proper theory of quantum gravity could be deduced from logical reasoning alone, for if uncertainty permeated all reasoning, as he implied, there could be no definite or certain starting point for logical reasoning. Eddington appears to recognize this limitation, since he introduced stabilization by assuming or taking for granted certain quantities (like the mass and charge of the electron), but his argument is *still* circular, because the mass of the electron, for example, is often found through experiments that involve Heisenberg's uncertainty principle in their analysis.

In dealing with this epistemological problem, Eddington notes in his *New Pathways in Science* that "we have been concerned to show that probability is always relative to knowledge (actual or presumed) and that there is no a priori probability of things in a metaphysical sense, i.e. a probability relative to complete ignorance."[32] Philosophically, this once again is a circular argument. For if probability is always relative to actual or presumed knowledge, then what is the first bit of knowledge, and how is it gleaned? Can not some probability be assigned to this first bit of knowledge? Eddington once again has locked himself in by assuming that there must be a first principle or quantity that serves as catalyst, but then attaching an uncertainty to it; or by assuming that the first principle or quantity arises through stabilization, but then violating his broad philosophical commitment to uncertainty. The more general problem here, philosophically speaking, is that mathematics is a deductive science that leaves little room to play with. Perhaps if Eddington had adopted a more physical approach in his work he might have been able to resolve this contradiction satisfactorily.

**The Aether**

Eddington's support of indeterminacy was not unique. Louis de Broglie (1892-1987) wrote in 1937: "In any case, in the present state of our knowledge, the Cartesian ideal of representing the physical world by means of 'figures and motion' seems to have suffered bankruptcy."[33]

Eddington quite likely thought about the concept of uncertainty early in his career, but the first clear evidence that he intended to introduce some level of uncertainty into the reference frame of a system is in a letter he wrote to Larmor in 1932.[34] Earlier, he had hinted that a particle was inseparable from its environment, but he had not discussed the uncertainty of the reference frame itself. Later, he developed this idea further to include as an origin of the reference frame the center of mass of an ensemble of particles,[35] and in his *Relativity Theory of Protons and Electrons* of 1936, he locates the origin in a Gaussian distribution of these particles,[36] which is how he treated it in his *Fundamental Theory*.[37]

Larmor and Eddington, who had exchanged letters for many years beginning in 1915,[38] had a curious relationship. As noted above, they disagreed about pacifism during the Great War,[39] but that seems only to have increased their respect for each other. Of most concern to us here are their respective views on the nature of the aether.

In 1900, Larmor (figure 2) published his well-known book, *Aether and Matter*, which Whittaker characterized as being in the classical mold.[40] Larmor believed in a "fluid" aether, that is, an aether as a tangible substance in (non-vacuum) space-time.[41] (Debate still goes on today regarding the nature of space-time and the "aether."[42]) Eddington's precise views about the nature of the aether are unclear, but he certainly assumed that it had some definite structure apart from the pure vacuum, which as we will see led him to his initial assumptions about uncertainty. In his 1932 letter to Larmor, he asked Larmor to examine some calculations he had made on an enclosed sheet of paper to describe the radiation emitted by a rotating ring of $n$ electrons.[43] He asked Larmor what would happen if one electron were removed from the ring, which then would become discontinuous as would the emitted radiation, since the "propagation of a discontinuity is a discontinuous process." He introduced a vector to describe the discontinuity, which he called the "aether displacement," and which linked the electron to the aether so that any measurement on the electron also required consideration of the aether. This idea later appeared in his concept of the uranoid (his name for the standard background environment), which is fundamental to his attribution of uncertainty to the reference frame.[44] He extended this idea to instantaneous states such as the present instant, "now": the "world-wide instant 'now' is created by ourselves and has no existence apart from our geocentric outlook...."[45] A four-dimensional world view removes such instantaneous states. Thus, uncertainty is inherent in space-time because the reference frame cannot be separated from the object under observation.

This relates to Eddington's picture of the aether, electricity, and gravitation. Gravitation is wedded to space-time: the curvature of space-time produces gravity. The aether then becomes the space-time background, which early modern cosmologists such as Eddington viewed as something tangible apart from the vacuum of space. Now consider the relationship of the aether to electricity. Electromagnetism was the first force to be combined with gravity in a single theory, first by Kaluza, then by Klein, Eddington, Einstein, and others. Eddington, in his letter to Larmor, thus considers electrical charge and the aether as inseparable and first expresses his concept of uncertainty in the reference frame through electromagnetism (see below).[46] He assigns the coordinates $\xi$, $\eta$, and $\zeta$ to the vector displacement of the aether in his derivation of the motion of the electron in its ring. Later, he carries this idea over in an early draft of his *Fundamental Theory*,[47] referring to any particle (charged and uncharged) as a "conceptual carrier" of these coordinates (finally described by Gaussian wave packets) and their conjugate momenta.

Eddington recognized that the aether was *not* matter, although it had substance. He made this point in another early draft of his *Fundamental Theory* when he asserted that to use the term "field" in place of "aether" was "ill-advised."[48] The strictly mechanical properties described both by second-rank tensors and wave mechanics allowed matter to look more like a field and was a simple way of describing its behavior. To Eddington, the aether thus was *not* matter.

**The Standard of Measurement**

The next major problem that Eddington had to confront both physically and philosophically stemmed from his conclusion that ordinary dimensional quantities cannot be used to define a unit of length such as the meter, because its physical nature constrained its accuracy. In his *Fundamental* Theory, Eddington held that pure numbers – and only pure numbers – could be used to define a quantitative unit of length.[49] By that time, the specification of a physical structure by pure numbers had already been developed in quantum theory (the number of elementary particles in a configuration state, for instance, was defined by a quantum number). Eddington thus concluded that the standard of length should be defined by a quantum-specified value, which is basically how it is defined today.

Eddington tackled this problem because he recognized the limitations of the system of measurement that was currently in use. In the 1930s the standard of length was the Paris Metre, which embodied an inherent uncertainty, as became evident in attempts to reproduce it elsewhere. Only in 1960 was an atomic standard adopted based on the wavelength of a particular red-orange spectral line emitted by krypton-86 in a gas discharge tube. Even here, however, the reproducibility of krypton-86 limited its accuracy. This standard thus was replaced in 1983 by the current one, which defines one meter as the length of the path traveled by light in a vacuum during 1/299,792,458 second. This implies that the speed of light is known exactly (which was not accepted universally in Eddington's time). This is a quantum-mechanical definition of just the type that Eddington wanted.

In its absence, Eddington defined his standard of length in terms of the periods of light waves and the amplitudes of vibrations of crystal lattices and found as a consequence that the speed of light was a constant. He had always believed this to be the case, but he now held it to be necessary to have an exact standard of measurement. He went so far as to make derogatory comments about those who held an opposing point of view.[50]

In specifying a quantum-mechanical system of measurement, Eddington also required the use of natural units of measurement. His system entailed that nothing could be measured only by a single unit (like that of length), but had to embody a connection between the gravitational constant $G$ and Planck's constant $h$. He asserted that a single unit of measurement (particularly a unit of length) meshed with his idea of uncertainty in the reference frame, and that the scale of a system actually was measured by an outside observer unless the system was the entire universe.[51] For a smaller system, an extraneous standard remained in it that had to be considered.

Eddington seems to have seen the fragmentation of the universe as being manifested in the different units of measurement, such as meters, grams, joules, and volts. How were the fragmented divisions tied together? He hinted that the simplest interpretation involves a single underlying phenomenon with an infinite number of manifestations embodying everything we can measure. This is quite profound philosophically. In Eddington's theory, a quantum-mechanical standard of length then would produce quantum-mechanical standards for every other measurable property in the universe, leaving only length to be calibrated.

**Observables and Coordinate Relations**

> It is well-known that the interference of different kinds of measurement is the source of Heisenberg's Uncertainty Principle which is the epistemological gateway by which the probability concept enters quantum theory.[52]

Clive W. Kilmister's words above serve to introduce the final merger of Eddington's ideas, which occurred when he found a common point of intersection between relativity and quantum mechanics. It centered initially on the concept of observables, later on that of coordinate systems. One of the early drafts of his *Fundamental Theory* has a nice description of observables.[53] He notes that in wave mechanics an observable is described by a product of two functions, while in relativity an observable is a relationship between two or more bodies (an observation point and observed object, or a reference point for a measurement between two observed objects).

In quantum-mechanical terms, the self-properties of two observables together with their conjugates, $\varphi^*\varphi$ and $\psi^*\psi$, are observationally equivalent to $\varphi^*\psi$. If a wave function represents a definite momentum of a particle, its position is entirely uncertain. In relativity these self-properties are represented by the stress-energy tensor $T_{\mu\nu}$. Eddington claimed that these self-properties represent a particle's *geometrical* coordinate system, while its position represents its *physical* coordinate system. Imagine a completely isolated particle at a point in outer space. According to Eddington, Heisenberg's uncertainly principle entails that the *geometrical* coordinates of this particle are uncertain. Now remove the particle and focus on the point itself. Eddington claimed that the location of this point is uncertain because it actually is the midpoint of a Gaussian probability distribution that represents the *physical* origin of the particle if it were returned to this point. Thus, Eddington believed that the very fabric of spacetime has an uncertainty in it. While Heisenberg took his uncertainty principle to refer to an uncertainty in the *geometrical* location of a particle, Eddington claimed that even if that uncertainty could be overcome, Heisenberg's uncertainty principle still would apply to its *physical* location and hence manifest itself. Thus, to Eddington, the physical fabric of spacetime has a probability or uncertainty attached to it.

This is one of Eddington's more esoteric concepts. We see that to develop his theory in detail, he was forced to distinguish between *geometrical* coordinates – the mapping of points onto a reference frame – and *physical* coordinates – the actual locations of objects in space. Statistical fluctuations in the actual physical reference frame then make uncertainty inherent in every measurement and preclude the possibility of overcoming Heisenberg's uncertainty principle. Thus, hidden variables are ruled out, because they would appear only in geometrical coordinates and not in the fluctuations of the reference frame itself. The curvature of space then "arises out of the statistical fluctuations of a distribution of a large number of particles."[54] Since that distribution had to obey Wolfgang Pauli's exclusion principle, gravity also was a consequence of this principle.

Eddington thus drew a "chain of connections" between relativity and quantum mechanics, which he called the "trunk road of relativistic quantum theory."[55] To this trunk road he attached representations of the geometrical and physical coordinates, which played a vital role in his concept of uncertainty in the reference frame. Thus, the relationship between these coordinates went beyond a statistical one and encompassed actual physical representations in both relativity and quantum mechanics independently.

**Against the Grain**

In another early draft of his *Fundamental Theory*, Eddington found that limiting wave mechanics to three dimensions introduced a factor of 4/5 into certain relationships. He attributed this anomaly to using known or stabilized masses instead of measured ones to simplify the solutions of his equations. This is essentially Kaluza-Klein in reverse – instead of extending the number of dimensions, Eddington *reduced* their number (or constrained the coordinates to act within a certain number of dimensions).

Eddington (figure 3) similarly went against the grain when he denounced the idea that a singularity could constitute a point of intersection between relativity and quantum mechanics and be treated as a pseudo-particle. (In 1931 Dirac had suggested the existence of magnetic monopoles,[56] which from time-to-time have played a role in theories of quantum gravity.) In quantum mechanics, a singularity is created by letting Planck's constant *h* go to zero. Eddington claimed that this was impossible, because the reciprocal of the fine-structure constant, $hc/2\pi e^2$, was an invariant and exactly equal to 137.[††] He insisted that the relativistic proper distance, when measured by a quantum-specified standard in Minkowski space-time, was the only link between quantum mechanics and gravity. He insisted on this point, yet repeatedly introduced curved space-time into his theory. Thus, he possibly viewed the universe as approximated by Minkowski space-time on a large scale, with curved space-time holding on a local scale, although he never said so explicitly. A group of researchers recently proved that the universe does have this geometry.[57]

In his *New Pathways in Science,* Eddington speculated on the origin of $\hbar$.[58] He noted that owing to Heisenberg's uncertainty principle the product of uncertainties in the position and momentum of a particle, for example, is on the order of Planck's constant *h*. In angular motion the greatest possible uncertainty is $2\pi$. Hence, the ratio $\hbar = h/2\pi$ accounts for the quantized orbits of electrons in atoms. Thus, Eddington viewed uncertainty as an inherent natural phenomenon, and it is not surprising that he extrapolated this concept beyond quantum mechanics to relativity theory.

**The Range Constant and Hubble Parameter**

---

[††] H.N. Russell quotes Eddington as saying in a cloakroom at the Stockholm meeting of the International Astronomical Union in 1938, "I always hang my hat on peg 137" (ref. 8, p. 135). An old story related to me by Laurie Brown at an American Physical Society meeting tells that Wolfgang Pauli (1900-1958) was also suspicious of this number and correctly predicted that he would die in the hospital after being assigned room 137.

Eddington culled two significant results from his work on uncertainty in his *Fundamental Theory* of 1946. The first was what he called the range constant *k* of nuclear (non-Coulombian) forces, which he found to be $k = 1.9 \times 10^{-13}$ centimeter. The earliest calculation of the range constant that I have found was carried out by George Gamow (1904-1968) in 1930 while developing his liquid-drop model of the nucleus.[59] Eddington, however, did not cite Gamow's paper; he noted instead that the range constant had been determined experimentally in proton-proton scattering experiments.

Eddington applied the range constant to an Einstein universe with significant consequences. He produced two equations with the radius $R_o$ of the universe and the number of particles *N* in the universe as the unknowns,[‡‡] which he solved simultaneously to find $N \sim 10^{79}$ and $R_o \sim 296$ megaparsecs (MPc).[60] Earlier, in his *New Pathways in Science* of 1935, he had derived the radius $R_o$ in a slightly different way based on the calculated mass of the universe.[61] His earliest calculation of the radius $R_o$, however, appeared in 1931,[62] where he found that $\frac{2\pi mc\alpha}{h} = \frac{\sqrt{N}}{R_0}$, where $\alpha = hc/2\pi e^2$. This differed from his 1935 result only by the presence of the constant $\alpha$, the reciprocal of the fine-structure constant, which he argued should be equal to 136 (in his *Fundamental Theory* he said 137[§§]). Simply substituting, however, we have $\frac{mc^2}{e^2} = \frac{\sqrt{N}}{R_0}$, where the right-hand side is the statistical variance of a large assemblage of particles in a volume of radius $R_o$, and the left-hand side is the ratio of the rest energy of a particle to the square of its charge. This result follows, Eddington said, since it "is well known that the ratio of $hc$ to $e^2$ is a pure number";[63] this thus represents an early attempt by him to explain the fine-structure constant, albeit a severely artificial one.

---

[‡‡] The two equations were $R_0 / N = Gm_H / \pi c^2 = 3.95 \times 10^{-53} \, cm$ and $R_0 / \sqrt{N} = k = 1.9 \times 10^{-13} \, cm$, where *G* is the gravitational constant and $m_H$ is the mass of a hydrogen atom.

[§§] A.V. Douglas (1895-1988) says that Eddington corrected this number to 137 in 1930 (ref. 9, p. 151), which he did, but he reverted to 136 in his 1931 paper before returning permanently to 137 in subsequent papers.

The limiting speed $V_0$ for the recession of the galaxies is given by $V_0 = \dfrac{c}{R_0\sqrt{3}}$, which, given his above value for the range constant $k$,*** he found to be $V_0 = 585$ km/s/MPc, whereas the observed value was 560 km/s/MPc. Using a new value for the range constant $k$, Eddington later refined this to be 572.4 km/s/MPc. These values appeared in his *Fundamental Theory* of 1946.[64] Had he lived to oversee its publication, however, he likely would have corrected them, since his earlier papers contain more accurate ones. For example, he gave 528 km/s/MPc in 1931,[65] at a time when both Edwin Hubble (1889-1953) and Willem de Sitter (1872-1934) had found 465 km/s/MPc, and in 1937 he gave 432 km/s/MPc.[66] He attributed the differences between his theory and observation to uncertainties in the distance scale and other astronomical problems, which he said introduced an error of close to 20 percent. Today $V_0$ is estimated to be about ten times smaller, around 50-70 km/s/MPc.

**Atomic Structure and the Philosophy of Free Information**

Another philosophical consequence of Eddington's use of uncertainty pertained to his theory of atomic structure, which again centered on his fascination with dimensionless

---

*** As seen in the next to last footnote, $R_o$ is related to the range constant $k$, and hence the recessional velocity $V_o$ also depends on $k$.

ratios.[†††] He noted that the ratio of the masses of the electron and proton could be described equally well as a ratio of their densities in the hydrogen atom, which is scale-free, as he explained in detail in his *Fundamental Theory*.[67]

Eddington expanded his theory to describe generalized systems of atoms whose coordinates are correlated rather than represented by wave functions. It might seem that atomic constituents should be represented by wave functions to achieve practical results, but correlating their coordinates renders this unnecessary. For instance, a proton and an electron in a closed vessel are equally likely to be anywhere in it at any given time; hence, based on the laws of probability, they eventually will combine to form a hydrogen atom with the emission of a photon. They still are equally likely to be anywhere in the vessel, but now their distribution functions are correlated. Thus, atomic wave functions are just correlated wave functions of atomic constituents and *not* distributed wave functions, a description that parallels the one that Erwin Schrödinger (1887-1961) employed in his first paper on wave mechanics of 1926. It also is reminiscent of density-functional theory, which is currently used in solid-state physics and quantum chemistry; it assumes that instead of all of an atom's information being packaged in a wave function, it is packaged in a density function. This method is often used today to calculate electron energy levels in atoms and molecules.

Atomic structure can be separated into two parts, one mechanical, the other electrical. Eddington referred to the former as scale-free, the latter as scale-fixed. To

---

[†††] Chandrasekhar wrote that Eddington never lost his fascination with large numbers, choosing frequently to write astronomical measures and distances with their zeros included explicitly. The number $136 \times 2^{256}$ is known as Eddington's number; it is the number of protons (with the same number of electrons) he believed to exist in the universe, and he wrote it out in full in his book, *The Philosophy of Physical Science* (ref. 50), p. 170. Chandrasekhar wrote in response: "Bertrand Russell asked Eddington if he had computed this number himself or if he had someone else do it for him. Eddington replied that he had done it himself during an Atlantic crossing!" (ref. 10, p. 3).

Eddington, mass and charge, the mechanical and electrical parts of an atom, are "free information" and lack uncertainty – there is no uncertainty, for instance, in the mass of the electron. This is debatable, of course, since the very act of measuring an atomic quantity, for instance in a collision process, introduces an uncertainty into the measurement. Thus, these atomic properties are not truly free information. Eddington was forced to accept this and thus to agree that an observer cannot really acquire any "free information." He therefore knew that to generalize and claim that the mechanical and electrical parts of an atom constitute "free information" went too far. He thus calls tabulated data, for instance values of the mass and charge of atomic particles, stabilized characteristics. Technically, in Eddington's theory, such stabilized characteristics are not observable, because all observations involve probability distributions as determined by experiment, while stabilized characteristics are part of the theory itself. In essence, Eddington thus argues that a stabilized characteristic must be derived from a theory and cannot be found experimentally—another of his controversial claims.

**Unification in a Nutshell**

I cannot go into Eddington's theory in detail mathematically; I can only state that his quest for a unified theory of quantum mechanics and relativity rested on a complex and profound treatment of uncertainty in the reference frame, as sketched above. He showed that the introduction of uncertainty into the reference frame entailed deep philosophical problems when dealing with observers and observables and prompted him to suggest several ideas that were unique at the time, including the concept of a quantum-mechanical standard of measurement similar to the one employed today. Some of his work also is reminiscent of modern techniques employed in quantum field theory. Physically, his theory separated the inertial and gravitational aspects of gravitation, while mathematically it dealt directly with uncertainty in the reference frame. This was the core of his posthumously published *Fundamental Theory*, but this work went through six drafts,[68] and had Eddington lived longer, he may have produced more. Thus, to grasp fully the evolution of Eddington's ideas, it is necessary to explore all of these drafts, some of which may contain insights that are significant today, just as Einstein's "greatest blunder" – his introduction of the cosmological constant $\Lambda$ – was found to be substantiated observationally in 1998. The cosmological constant also played an important role in Eddington's *Fundamental Theory*; he believed it to be non-zero: "To set $\Lambda = 0$ is to knock the bottom out of space."[69]

**Genius Unwrapped**

Eddington was a consummate and meticulous mathematician. Even his handwriting is extremely neat and compact. He defended his points of view passionately, even aggressively, as shown in his letters to Larmor and in some of his published works. He often seems to denigrate his opposition, but never seems to have held a grudge. His early rocky relationship with Larmor ended in friendship, to the point that he helped to organize Larmor's 70$^{th}$-birthday gala in 1927.[70] He and Milne (figure 4) also criticized each other's scientific views harshly, but were close personal friends. For example, when Eddington reviewed one of Milne's books, he wrote to Milne, saying: "I realise that the review can scarcely be pleasing to you; but I hope you will recognise that it might have been worse if (holding the opinion I do) I had let myself go without regard to our friendship."[71] He ended his letter by chatting wistfully about an eight-day bicycle tour he intended to begin the following day, planning to let "the worries of the universe fade into their proper significance." Eddington's solitary cycling tours in the spring and fall of each year were well-known among his friends. It was less well-known that he kept careful records of them.[72]

Eddington always searched for physical justifications for his mathematics, despite its deductive nature. Moreover, his philosophy of science was far from unusual at the time, although in his later years his contemporaries often viewed him as a recluse and even a renegade. His work was regarded as heterodoxical, both philosophically and scientifically.[73] It was received unfavorably particularly in Britain, where his harshest critic was the astrophysicist and philosopher Herbert Dingle, who referred to the theories of Eddington, Milne, Dirac, and others as the "pseudoscience of invertebrate cosmythology."[74] Eddington's mathematics, however, were fairly standard, and he always took pains to ensure their rigor. Thus, the charge of heterodoxy generally focuses on his physics. But his was hardly the only physics before or since that was regarded as heterodoxical. Dirac's hole theory is another case in point, although even today it appears in textbooks as a pedagogical tool.[75]

Eddington employed a number of methods that are in use today, including the introduction of chirality into particle physics, and successfully found some important results, including his independent discovery of Majorana spinors.[76] He recognized the need for a quantum-mechanical standard of length, and he advocated the inseparability of an object and its environment, a philosophical stance that appears in some versions of quantum field theory. His idea that there is a fundamental link between quantum mechanics and relativity based on the concept of coordinates also is close to our modern view: Both theories rely heavily on coordinates, and both employ topological concepts. Apart from his pioneering work on cosmology, he almost single-handedly founded the field of stellar structure. About those critics who laughed at Eddington, Einstein once said: "Why should they laugh? They have never done what he has done!"[77] In any case, my analysis of Eddington's work on uncertainty in the reference frame offers a glimpse into the mind of one of the greatest mathematical astronomers of the twentieth century.

Eddington died following an operation for cancer on November 22, 1944, in the Evelyn Nursing Home in Cambridge. The Astronomical Society of the Pacific deemed his death a "calamity."[78] Had he lived longer, he may have provided answers to some of the questions he left hanging, but more likely he would have raised yet more questions.

**Acknowledgements**

I gratefully acknowledge the support and encouragement of Edmund Robertson, John O'Connor, and the School of Mathematics and Statistics at the University of St. Andrews and of Velda Goldberg and the Department of Physics at Simmons College. I thank the staff of the University of St. Andrews library for their assistance, and the staff of the Royal Society's Library and Sackler Archive who gave me access to the correspondence in their collections. I thank Peter Hingley at the Royal Astronomical Society's library for supplying the photographs of Eddington and for suggesting the one where he is speaking at the Kepler memorial in Heidelberg. I am grateful to the Simmons College Library for obtaining hard-to-find books for me on inter-library loan and for providing me access to hundreds of journal articles through their subscription to JSTOR. I thank the Massachusetts Institute of Technology for providing me access to their library and Retrospective Collection (RSC) that contains hard-to-find books. I thank Len Soltzberg of Simmons College for useful conversations on the history of quantum mechanics, Alex Craik of the University of St. Andrews for discussing the general subject of my paper with me, John Amson for a useful discussion on post-Eddingtonian theory, and Roger H. Stuewer for his careful editorial work on my paper. Finally, I thank Mrs. Meg Weston Smith, the daughter of E.A. Milne, whose generosity, warm hospitality, and openness have made my project most enjoyable. Her knowledge of her father's work and friendship with Eddington, and of the large collection of her father's materials that she maintains, were invaluable to me in my work.

**References**

[1] Theodor Kaluza, "Zum Unitätsproblem der Physik," *Sitzungsberichte der Preussische Akademie der Wissenschaften* (1921), 966-972. Kaluza sent this paper to Einstein in 1919, but Einstein did not communicate it for publication until 1921.

[2] Oskar Klein, "The atomicity of electricity as a quantum theory law," *Nature* **118** (1926), 516.

[3] P.A.M. Dirac, "The Quantum Theory of the Electron," *Proceedings of the Royal Society of London* [A] **117** (1928), 610-624; "The Quantum Theory of the Electron, Part II," *ibid*., [A] **118** (1928), 351-361. For a full discussion, see Helge Kragh, *Dirac: A Scientific Biography* (Cambridge: Cambridge University Press, 1990), Chapter 3, pp. 48-66.

[4] Arthur S. Eddington, "A Symmetrical Treatment of the Wave Equation," *Proc. Roy. Soc*. [A] **121** (1929), 524-542.

[5] Arthur S. Eddington, *New Pathways in Science* (New York: Macmillan, 1935), p. 84.

[6] P.A.M. Dirac, "The Cosmological Constants," *Nature* **139** (1937), 323; for a discussion, see Kragh, *Dirac* (ref. 3), pp. 228-229.

[7] Arthur S. Eddington, *Fundamental Theory* (Cambridge: Cambridge University Press, 1946).

[8] Henry Norris Russell, "Arthur Stanley Eddington, 1882-1944," *Astrophysical Journal* **101** (1945), 135.

[9] Allie Vibert Douglas, *The Life of Arthur Stanley Eddington* (London: Thomas Nelson, 1956), p. 103.

[10] Subramanyan Chandrasekhar, *Eddington: The most distinguished astrophysicist of his time* (Cambridge: Cambridge University Press, 1983), p. 25.

[11] *Ibid*., p. 25.

[12] Eddington to Larmor, June 7, 1916, Royal Society Archive.

[13] For example, see Eddington to Milne, April 8, 1935, in possession of Mrs. Meg Weston Smith.

[14] Douglas, *Eddington* (ref. 9), p. 2; H.C. Plummer, "Arthur Stanley Eddington 1882-1944," *Royal Society of London Obituary Notices* **5** (1945-1948), 114.

[15] W.M. Smart, "Sir Arthur Stanley Eddington, O.M., F.R.S.," *The Observatory* **66** (1945), 1; Plummer, "Eddington" (ref. 14), pp. 113-114.

[16] Eddington to Schuster, 1909 (month and day not noted), Royal Society Archive.

[17] Smart, "Eddington" (ref. 15), p. 2.

[18] Eddington to Milne, April 8, 1935 (ref. 13).

[19] Chandrasekhar, *Eddington* (ref. 10), p. 3.

[20] Eddington, *New Pathways* (ref. 5), p. 62.

[21] H. Spencer Jones and Edmund Whittaker, obituary, *Mon. Not. Roy. Ast. Soc.* **105** (1945), 75.

[22] Chandrasekhar, *Eddington* (ref. 10), p. 6.

[23] Eddington, *New Pathways* (ref. 5), Chapters IV-VI, pp. 72-134.

[24] *Ibid*., pp. 124-125.

[25] Joan Richards, "The Probable and the Possible in Early Victorian England," in Bernard Lightman, ed., *Victorian Science in Context* (Chicago: University of Chicago Press, 1997), p. 59.

[26] Arthur S. Eddington, "The Envelopes of Comet Morehouse," *Monthly Notices of the Royal Astronomical Society* **70** (1910), 442-458.

[27] Arthur S. Eddington, "The Total Eclipse of 1919 May 29 and the Influence of Gravity on Light," *The Observatory* **42** (1919), 119-122.

[28] J.E. Warriner, Warriner's English Grammar and Composition, (Harcourt Brace Jovanovich, Orlando, 1986).

[29] Richards, "The Probable and the Possible" (ref. 25), p. 59.

[30] Bertrand Russell, *Introduction to Mathematical Philosophy* (London: George Allen and Unwin, 1919), pp. 3-4.

[31] *Ibid*., p. 7.

[32] Eddington, *New Pathways* (ref. 5), p. 133.

[33] Louis de Broglie, *Matter and Light* (New York: Norton, 1937), p. 247.

[34] Eddington to Larmor, October 24, 1932, Royal Society Archive.

[35] Clive William Kilmister, *The Fundamental Theory: Eddington's search for meaning* (Cambridge: Cambridge University Press, 1994), p. 228.

[36] Arthur S. Eddington, *Relativity Theory of Protons and Electrons* (Cambridge: Cambridge University Press, 1936), p. 107.

[37] Eddington, *Fundamental Theory* (ref. 7), p. 3.

[38] Eddington to Larmor, January 23, 1915, Royal Society Archive.

[39] Eddington to Larmor, June 7, 1916 (ref. 12).

[40] Edmund Whittaker, *A History of the Theories of Aether & Electricity.* Vol. 1. *The Classical Theories* (New York: Harper, 1951), p. 303.

[41] *Ibid.*, pp. 287-288; Helge Kragh, *Quantum Generations: A History of Physics in the Twentieth Century* (Princeton: Princeton University Press, 1999), p. 89.

[42] Frank Wilczek, "The Persistence of Ether," *Physics Today* **52** (January 1999), 11.

[43] Eddington to Larmor, October 24, 1932 (ref. 34).

[44] Eddington, *Fundamental Theory* (ref. 7), pp. 13-14.

[45] Arthur S. Eddington, *The Theory of Relativity and its Influence on Scientific Thought* (Oxford: Clarendon Press, 1922), p. 17.

[46] Eddington to Larmor, October 24, 1932 (ref. 34).

[47] Noel Bryan Slater, *The Development and Meaning of Eddington's 'Fundamental Theory' Including a Compilation from Eddington's Unpublished Manuscripts* (Cambridge: Cambridge University Press, Cambridge, 1957), p. 60.

[48] *Ibid.*, p. 72.

[49] Eddington, *Fundamental Theory* (ref. 7), p. 7.

[50] Eddington, *The Philosophy of Physical Science* (Cambridge: Cambridge University Press, 1939), pp. 78-79.

[51] Eddington, *Fundamental Theory* (ref. 7), p. 14.

[52] Kilmister, *Fundamental Theory: Eddington's* (ref. 35 ), pp. 98-99.

[53] Slater, *Development and Meaning* (ref. 47), p. 68.

[54] *Ibid.*, p. 73.

[55] *Ibid.*, p. 81.

[56] P.A.M. Dirac, "Quantized Singularities in the Electromagnetic Field," *Proc. Roy. Soc.* [A] **133** (1931), 60-72; for a discussion, see Kragh, *Dirac* (ref. 3), pp. 209-214.

[57] P. de Bernardis, et al., "A Flat Universe from High-resolution Maps of the Cosmic Microwave Background Radiation," *Nature*, **404** (2000), p955.

[58] Eddington, *New Pathways* (ref. 5), p. 105.

[59] George Gamow, "Mass Defect Curve and Nuclear Constitution," *Proc. Roy. Soc.* [A] **126** (1930), 632-644, especially 635. For a full treatment, see Roger H. Stuewer, "The Origin of the Liquid-Drop Model and the Interpretation of Nuclear Fission, *Perspectives on Science* **2** (1994), 76-129, especially 78-87.

[60] Eddington, *Fundamental Theory* (ref. 7), p. 10.

[61] Eddington, *New Pathways* (ref. 5), pp. 248-250.

[62] Arthur S. Eddington, "On the Value of the Cosmical Constant," *Proc. Roy. Soc.* [A] **133** (1931), 605-615.

[63] Arthur S. Eddington, "The Charge of an Electron," *Proc. Roy. Soc.* [A] **122** (1929), 358-369, especially 358.

[64] Eddington, *Fundamental Theory* (ref. 7), p. 10.

[65] Eddington, "Value of the Cosmical Constant" (ref. 61), p. 614.

[66] Eddington, "The Cosmical Constant and the Recession of the Nebulae," *American Journal of Mathematics* **59** (1937), 1-8, especially 7.
[67] Eddington, *Fundamental Theory* (ref. 7), p. 18; see also Slater, *Development and Meaning* (ref. 47), p. 89.
[68] Douglas, *Eddington* (ref. 9), p. 183.
[69] Chandrasekhar, *Eddington* (ref. 10), p. 39.
[70] Eddington to Larmor, April 13, 21, 22, 1927, Royal Society Archive.
[71] Eddington to Milne, April 8, 1935 (ref. 13).
[72] Chandrasekhar, *Eddington* (ref. 10), p. 5.
[73] Kragh, *Quantum Generations* (ref. 41), pp. 218-229.
[74] Quoted in *ibid.*, p. 227.
[75] For example, see Richard L. Liboff, *Introductory Quantum Mechanics* (Reading, Mass.: Addison-Wesley, 1998), pp. 828-829.
[76] Chandrasekhar, *Eddington* (ref. 10), pp. 54-55.
[77] Quoted in Douglas, *Eddington* (ref. 9), p. 146.
[78] Anonymous, "Arthur Stanley Eddington," *Publications of the Astronomical Society of the Pacific* **57** (1945), 116.

Mathematical Institute

University of St. Andrews

North Haugh, St. Andrews, Fife, KY16 9SS Scotland

e-mail: ian@dcs.st-and.ac.uk

Department of Physics

Simmons College

300 The Fenway

Boston, MA 02115 USA

e-mail: durhami@simmons.edu